\documentclass[nopreprintline]{elsarticle}
\usepackage[margin=3cm]{geometry}

\usepackage{amssymb}
\usepackage{caption}
\usepackage{multirow}
\usepackage{upgreek}
\usepackage{bm}
\usepackage{tabto}
\usepackage{subcaption}
\usepackage{natbib}
\setcitestyle{authoryear,open={(},close={)}} 

\journal{}

\begin{document}

\begin{frontmatter}

\title{COVID-19 detection using chest X-rays: is lung segmentation important for generalization?}

\author{Pedro R. A. S. Bassi*\textsuperscript{1}\textsuperscript{2}\textsuperscript{3}, Romis Attux\textsuperscript{1}}

\address{\textsuperscript{1}Department of Computer Engineering and Industrial Automation, School of Electrical and Computer Engineering, University of Campinas - UNICAMP. 13083-970, Campinas, SP, Brazil. 

\textsuperscript{2}Current: Alma Mater Studiorum - University of Bologna. 40126, Bologna, BO, Italy.  

\textsuperscript{3}Current: Istituto Italiano di Tecnologia. 16163, Genova, GE, Italy.

*E-Mail: pedro.salvadorbassi2@unibo.it.}

\begin{abstract}

\emph{Purpose:} We evaluated the generalization capability of deep neural networks (DNNs), trained to classify chest X-rays as Covid-19, normal or pneumonia, using a relatively small and mixed dataset. \emph{Methods:} We proposed a DNN to perform lung segmentation and classification, stacking a segmentation module (U-Net), an original intermediate module and a classification module (DenseNet201). To evaluate generalization, we tested the DNN with an external dataset (from distinct localities) and used Bayesian inference to estimate probability distributions of performance metrics. \emph{Results:} Our DNN achieved 0.917 AUC on the external test dataset, and a DenseNet without segmentation, 0.906. Bayesian inference indicated mean accuracy of 76.1\% and [0.695, 0.826] 95\% HDI (highest density interval, which concentrates 95\% of the metric's probability mass) with segmentation and, without segmentation, 71.7\% and [0.646, 0.786]. \emph{Conclusion:} Employing a novel DNN evaluation technique, which uses LRP and Brixia scores, we discovered that areas where radiologists found strong Covid-19 symptoms are the most important for the stacked DNN classification. External validation showed smaller accuracies than internal, indicating difficulty in generalization, which is positively affected by segmentation. Finally, the performance in the external dataset and the analysis with LRP suggest that DNNs can be trained in small and mixed datasets and still successfully detect Covid-19.

\end{abstract}


\end{frontmatter}


\section{Introduction}

Diagnosis is an important aspect for controlling Covid-19 spread and helping infected patients. Nowadays, an active SARS-CoV-2 infection is normally diagnosed with the detection of its RNA (ribonucleic acid) genome (usually employing reverse transcription–quantitative polymerase chain reaction, RT–qPCR, or, alternatively, using next-generation sequencing, NGS, or isothermal nucleic acid amplification assays) or with antigen tests, which assess the presence of viral proteins (\cite{diagnosis}). Having higher sensitivity (50-70\% in a real clinical scenario) and specificity (~99\%), to this date (August 2022), the gold-standard Covid-19 diagnosis method is RT-qPCR (\cite{test},\cite{diagnosis}). However, this method is expensive, requires a considerable amount of time, and is at high demand during infection peaks. 

X-ray is one of the cheapest and most available Covid-19 alternative detection methods, mostly when the disease spread in developing countries is considered. Covid-19 has characteristic signs which can be observed in chest X-rays, like bilateral radiographic abnormalities, ground-glass opacity, and interstitial abnormalities (\cite{clinicalCOVID}). However, the images' analysis can be complicated. Thus, we think that artificial intelligence will be able to help in the creation of a reliable system to help clinicians in this task.

Deep neural networks (DNN) for Covid-19 detection were already proposed (\cite{reviewCovid}). However, some researchers raised concerns about the possibility of bias. \cite{ArtigoCritico} mixed different chest X-ray datasets, removed most of the lungs from the images and trained DNNs to classify to which dataset the images belonged. They were able to obtain high accuracies and, according to them, this reveals that dataset biases may influence DNNs trained with mixed datasets, reducing their generalization capability. We do not think this test alone shows that the biases are strong enough to highly influence DNN decisions (a DNN is a very flexible model: if the relevant information in the X-rays is deleted, it may be able to learn even very small dataset particularities). But we agree that the study proves the existence of dataset bias.

Accordingly, a review (\cite{ShortcutCovid}) has shown that if the DNNs are allowed to analyze the entire X-ray, they tend to focus on areas outside of the lungs. The study suggested that the DNNs pay attention to X-ray features that are not representative of the disease symptoms (like text outside of the lungs), i.e., they focus on image characteristics that represent bias. Analyzing these features, the DNN can achieve high accuracy on the training dataset and standard test databases, which are independent and identically distributed (i.i.d.) in relation to the training data (as they also present the features). However, the DNNs do not properly generalize to real-world scenarios or out-of-distribution (o.o.d.) datasets, whose X-rays are gathered from external sources in relation to the training samples. This phenomenon is known as shortcut learning, and the review shows that, in Covid-19 detection, performances on i.i.d. test databases can be unrealistically high (\cite{ShortcutCovid}). 

In March 2021, we still cannot find an open and large Covid-19 X-ray dataset, with all images collected from the same sources. This would be the best case scenario, as different classes would not present different biases. But Covid-19 classification datasets are generally relatively small and mixed, i.e., different classes have different sources (\cite{reviewCovid}). Our objective is to understand how, in a dataset like this, bias affects a DNN classifying healthy individuals, Covid-19 and pneumonia (which is a disease that also creates abnormalities in chest X-rays, such as airspace consolidation, poorly defined small centrilobular nodules, and bilateral asymmetric ground-glass opacity, \cite{clinicalPneumonia}). Therefore, we used external testing and validation (holdout) databases, whose X-rays were not from the hospitals that provided the training images. We can also refer to the external datasets as out-of-distribution (o.o.d.) in relation to the training database. Furthermore, we analyzed if the utilization of lung segmentation can improve performance on the external test dataset, which would indicate a reduction of bias and improved generalization.

In this study we use a large DNN, which performs lung segmentation, and then classifies the segmented images. We trained for classification with twice transfer learning, downloading ImageNet (\cite{imagenet}) pretrained classification networks, training them on a large lung disease classification database (\cite{NIHSet}) and then on our dataset (including Covid-19, normal and pneumonia). 

We evaluated our networks with traditional performance measurements (point estimates). But, due to the small number of available Covid-19 X-rays, our test dataset is small. This lowers the reliability of these measurements when predicting the classifier real-world performance. Therefore, we quantified the measurements' uncertainty, using a Bayesian model (\cite{bayesianEstimator}) to estimate the performance metrics probability distributions and their statistics, like 95\% highest density intervals (an interval containing 95\% of the metric probability mass, and whose points have probabilities that are higher than any point outside of it). We expanded the model in \cite{bayesianEstimator} to also estimate class specificity and mean specificity. 

We employed a technique called Layer-wise Relevance Propagation or LRP (LRP \cite{LRP}) to create heatmaps of the X-rays, showing which areas most contributed to the classification and which were more representative of other classes. These maps help us to better understand how DNNs make decisions, improving interpretability. They also show if the proposed DNN is truly ignoring the unimportant information outside the lungs and allow us to compare how the two trained models are classifying the images. Finally, the maps may be helpful for a clinician in finding the Covid-19 signs in an X-ray and evaluating the DNN prediction. 

This study presents a large DNN, containing 3 stacked modules. The segmentation module is an U-Net (\cite{unet}), trained beforehand to receive X-rays and output segmentation masks (which are white in the lung regions and black everywhere else). Then comes an original intermediate module, which uses the U-Net output and the input image to erase the unimportant X-rays regions, and performs batch normalization. Finally, the classification module, a 201-layers dense neural network (\cite{DenseNet}), returns the probabilities of the X-ray containing healthy lungs, pneumonia or Covid-19. We compare this network to a DenseNet201.

This work introduces a new technique to compare DNN's analysis of Covid-19 X-rays to radiologists', using LRP and X-rays scored with the Brixia scoring system, which is a methodology created for radiologists to semi-quantitatively score Covid-19 severity in six lung zones (\cite{brixia}). Please refer to Section \ref{brixSec} for a detailed explanation of the scoring system.  Do they look at the same Covid-19 signs? Is there a correlation between areas where radiologists find more severe symptoms to areas with more relevance in heatmaps? Do DNNs predict higher Covid-19 probabilities in X-rays with higher Brixia scores?

The main contribution of this paper consists in a profound analysis of the effects of lung segmentation on generalization in the field of Covid-19, using a test dataset created by external sources (with respect to the training dataset). Novel aspects of this analysis are the utilization of Bayesian inference to estimate the performance metrics probability distributions and the comparison of LRP heatmaps with X-rays analyzed using the Brixia score. Finally, we suggested a modular DNN architecture, composed of two state-of-the-art DNNs and an original intermediate module. It is possible to utilize just our trained segmentation module, along with the intermediate one, attach it to an alternative classification network and train for classification. This can provide a simple and fast way to create other DNNs that perform segmentation and classification. Our trained DNNs are available for download at https://github.com/PedroRASB/Covid-19-detection-with-lung-segmentation.

\section{Methods}

In this section we explain the employed datasets, the data processing and augmentation procedures, the deep neural networks and their training schemes, and, finally, the LRP procedure and Bayesian model used to analyze this study’s results. 

\subsection{The source databases}
In this section, we describe the databases we utilized to create the datasets that we used in this study. 

\subsubsection{NIH ChestX-ray14 dataset (\cite{NIHSet})}
ChestX-ray14 is a very large dataset of frontal chest X-rays, containing 112120 images, from 30805 patients, showing 14 different thoracic diseases, as well as healthy individuals. The dataset was originally created by the US National Institutes of Health and the authors automatically labeled it with Natural Language Processing, using radiological reports. The labels have an estimated accuracy that is higher than 90\% (\cite{NIHSet}).

It is an unbalanced dataset and a single patient can have more than one disease, therefore, classifying the database is a multi-label classification problem. The dense neural network CheXNet (\cite{chexnet}) was trained on this dataset.

925 images, showing healthy patients, were extracted from this database and used in our classification training dataset. Those images correspond to 925 different patients, with mean age of 46.8 years (with 15.6 years of standard deviation) and who are 54.3\% male. Additionally, 1295 ChestX-ray14 images, showing patients with pneumonia, were also included in our classification training dataset. They correspond to 941 patients, with a mean age of 48 years (standard deviation of 15.5 years), and who are 58.7\% male.

\subsubsection{Montgomery and Shenzen datasets (\cite{ChineseDataset2})}
This database was created by the National Library of Medicine, National Institutes of Health, Bethesda, Maryland, USA, in collaboration with the Department of Health and Human Services, Montgomery County, Maryland, USA and with Shenzhen No.3 People's Hospital, Guangdong Medical College, Shenzhen, China (\cite{ChineseDataset2}). The X-rays taken in Shenzen show 336 normal cases and 326 tuberculosis cases. In the Montgomery images there are 80 normal cases and 58 tuberculosis cases.

The Montgomery images also came with segmentation masks, created under the supervision of a radiologist (\cite{ChineseDataset1}, \cite{ChineseDataset2}). The dataset authors segmented the images excluding the lung part behind the heart, and following some anatomical landmarks, such as the ribs, the heart boundary, aortic arc, pericardium line and diaphragm (\cite{ChineseDataset1}, \cite{ChineseDataset2}).

The authors in \cite{ShenMasks} created segmentation masks for most of the Shenzen database. They are similar to the Montgomery's (e.g., they also exclude the lung part behind the heart).

The healthy patients in the Montgomery and Shenzen database have a mean age of 36.1 years (with standard deviation of 12.3 years) and are 61.9\% male. Their X-rays were used in our classification training dataset.

\subsubsection{Covid-19 dataset (\cite{GitCovidSet})}

Covid-19 image data collection (\cite{GitCovidSet}) is one of the largest Covid-19 X-ray databases to date.
The dataset also contains other kinds of pneumonia, including viral variants (such as Middle East respiratory syndrome/MERS, and severe acute respiratory syndrome/SARS) and bacterial pneumonia, but we did not use them in this study. From this dataset, we obtained 475 Covid-19 X-rays (all the frontal Covid-19 X-rays).

It is a public open dataset, whose images were collected from public sources or indirectly from hospitals and clinicians (\cite{GitCovidSet}). It is the largest public collection of Covid-19 chest X-rays we found, and is also well documented. For example, it contains information about patient age, gender and the image source.

The images we utilized correspond to 295 Covid-19 patients, with a mean age of 42.5 years (with standard deviation of 16.5 years) and who are 64.5\% male. We have information about disease severity on some of them: from 87 patients, 79.3\% survived; from 118 patients, 61.9\% needed ICU; from 77 patients, 61\% were intubated; from 107 patients, 62.6\% needed supplemental oxygen.

\subsubsection{CheXPert dataset (\cite{irvin2019chexpert})}
The CheXPert database contains images from the Stanford University Hospital. It has 224313 chest X-rays, from 65240 patients, showing 13 thoracic diseases or no findings (\cite{irvin2019chexpert}). As in the NIH ChestX-ray14 dataset, the images were automatically labeled, by the database authors, using Natural Language Processing to analyze radiological reports. The labels' estimated accuracy is also above 90\%. The exceptions, in the images we used, are 8 pneumonia X-rays and 26 normal X-rays, which were manually labeled by three board certified radiologists (these images are part of the original CheXPert test dataset).

We used part of the CheXPert database in our classification dataset, as part of the external validation. 79 pneumonia and 79 healthy images were used, including the ones manually labeled by three radiologists. The normal images correspond to 73 patients, with a mean age of 49.5 years (with standard deviation of 18.5 years) and who are 56.2\% male. The pneumonia images correspond to 61 patients, with mean age of 61.9 years (standard deviation of 18.1 years), and who are 60.7\% male.

\subsection{The segmentation dataset}
This dataset was used to train an U-Net to segment the lungs in frontal chest X-ray images. It contains images of Covid-19 (327), pneumonia (327), normal lungs (327) and tuberculosis (282). Pediatric patients were excluded from this study, because the Covid-19 database youngest patient is 20 years old and we thought that adding children to the other classes could create bias during classification (training the DNN not to associate children with Covid-19).

The normal and tuberculosis images were all the X-rays in Montgomery and Shenzen datasets that had corresponding segmentation masks; thus, in the segmentation dataset we excluded the normal and tuberculosis X-rays without segmentation targets. The pneumonia X-rays were randomly selected from the NIH ChestX-ray14 images. The Covid-19 images were randomly taken from the Covid-19 database (\cite{GitCovidSet}).

As targets, this dataset contains segmentation masks for each X-ray. For the healthy and tuberculosis images the masks were already provided in the Montgomery database and in \cite{ShenMasks}, for the Shenzen dataset. We created the other segmentation masks (for pneumonia and Covid-19). The mask creation process will be described with more details in Section \ref{masksSection}.

\subsubsection{Segmentation dataset subdivisions}
We separated the segmentation dataset in 3: training, validation and testing. We used them to train our U-Net with hold-out validation. The dataset subdivisions were random, but we performed a patient split: if we had more than one image from the same patient, all of them were used in only one subdivision.

For testing we used 150 images, 50 from each class (pneumonia, Covid-19 and normal, with 10 from Montgomery and 40 from Shenzen). We did not include tuberculosis images here because this class is not present in our classification dataset, thus the U-Net performance on it was not as relevant. But they were included in training and validation because we thought that more images would generate a better segmentation neural network.

To create the training and validation datasets we removed the test images, then randomly selected 80\% of the remaining X-rays as training and 20\% as validation. We kept both datasets balanced.

\subsection{Classification training dataset}
We used this dataset to classify chest X-ray images in one of three classes: healthy, pneumonia or Covid-19. It consists of frontal X-rays and has 1295 images of healthy subjects, 1295 of pneumonia patients and 396 of Covid-19 patients. Unlike the segmentation dataset, which had masks, this dataset has simple classification labels: Covid-19, normal or pneumonia.

The coronavirus images were all Covid-19 frontal X-rays in \cite{GitCovidSet}, except for the ones from Hannover Medical School, Hannover, Germany (they will be used in the external testing and validation datasets). The pneumonia X-rays were NIH ChestX-ray14 images labeled as pneumonia and with adult patients. Finally, the healthy images were all normal images from the Montgomery and Shenzen databases (with adult patients), along 925 normal images from  ChestX-ray14 (randomly selected, among adults). 

\subsection{External classification dataset}
We used the external classification dataset for validation (holdout) and testing when training for Covid-19 detection. 

We did not get the external Covid-19 images from another coronavirus database, because, as the current availability of Covid-19 X-rays is still limited, different datasets can have the same images. Instead, we separated the Covid-19 image data collection (\cite{GitCovidSet}) in two, according to geographical location. We chose all the images from Hannover Medical School (Hannover, Germany) for the external dataset because there are 79 images from this locality, a reasonable amount to create a validation and a test dataset (considering the small number of Covid-19 images), and because there are only 3 other images from Germany in the entire dataset (from Essen and Berlin). Therefore, the chance of a patient from Hannover having X-rays in another hospital from our database is very small.

The images for the normal and pneumonia classes were extracted from the CheXPert database. 79 images from each class were randomly selected, among the adult patients. We included, in the external dataset, all the normal and pneumonia images labeled by the three radiologists.

We divided the external dataset in two, for test and validation. The test dataset included 50 images from each class, and the validation dataset, 29. The division was random, but we did not allow X-rays from a single patient to be in more than one dataset.

\subsection {Image preprocessing}
Original image sizes varied between datasets or sometimes even within the same dataset, and we decided to utilize the input shape of 224x224, with 3 channels. This is the DenseNet original input size, and also the shape that we successfully used in our previous work with Covid-19 detection in X-rays (\cite{bassi2021deep}). With 3 channels we can take better advantage of transfer learning, due to the convolutional kernel shapes; images larger than 224x224 would be more detailed, but they would cause the simulations to be much slower, and a large input shape with a small training dataset can make the data very sparse in the input space, aggravating the problem of overfitting (\cite{Curse}). Therefore, although we think that the exploration of different input shapes is an important research topic in the context of Covid-19 detection, we used the ImageNet standard of 224x224, because this choice had already been successful (\cite{bassi2021deep}) and because it does not detract from the main purpose of this paper, which is to analyze generalization on an external dataset and the effects of lung segmentation.

When we loaded the X-rays, we converted them to grayscale and single-channel images (using OpenCV), with pixel values ranging from 0 to 255. We did this in order to remove any color information from the datasets, because some images had slight color variations, which could become a source of bias. As the DenseNet original input size is 224x224 with 3 channels, we converted the grayscale images to RGB (replicating the single-channel pixel values into three channels). Afterwards, we applied histogram equalization and normalized the pixel values between 0 and 1. Finally, we resized the X-rays to 224x224.

In the external test and validation datasets, as well as the segmentation datasets, we made the images square (if they were not already) by adding black bars in their borders, before resizing. We used the black bars to avoid changing the X-rays aspect ratio. Furthermore, as we did not use the bars in the classification training dataset, the DNNs (especially the one without segmentation) could not learn to identify them.

\subsection{Training for segmentation}

\subsubsection{The U-Net}
The U-Net architecture was proposed in \cite{unet}, as a DNN for segmentation in biomedical databases. Therefore, It was designed to perform well using a small quantity of annotated samples and a large amount of data augmentation. For example, the authors in \cite{unet} used the DNN to segment neuronal structures in electron microscopic stacks, winning the International Symposium on Biomedical Imaging (ISBI) cell tracking challenge in 2015. As we had a relatively small amount of lung X-rays with masks, the U-Net seemed like a good option for lung segmentation.

The architecture was already used for this purpose. In \cite{TuberculosisUNet} the authors used an U-Net to successfully segment lungs in chest X-rays, generating masks that were used to create a new dataset, with images that contained only the lungs (and black pixels outside them). Afterwards, they classified these images as tuberculosis or non-tuberculosis, utilizing convolutional neural networks (CNNs).

An U-Net is a fully convolutional DNN with two symmetric paths, a contracting path, which captures context in the image, and an expanding path, which allows precise localization. The paths are connected in multiple points. More information can be found in \cite{unet}.

Our U-Net implementation is the same as the original (shown in Figure 1 of \cite{unet}), it has 5 blocks in each path, each one with two 2D convolutions and ReLu activation.

\subsubsection{Training with the Montgomery and Shenzen databases}
We trained an U-Net with the Shenzen and Montgomery datasets, using their manually created segmentation masks as targets. We randomly selected 70\% of the images for training, 20\% for validation (hold-out) and 10\% for testing. We used data augmentation in the training dataset, multiplying the number of images by 8 (the original images were not removed), with random rotations (between -40 and 40 degrees), translations (with a maximum of 28 pixels up or down and also 28 left or right) and horizontal flipping (50\% chance). 

During every training procedure in this work, we used the validation error to estimate the DNN with the best generalization capability, and this network was then evaluated on the test dataset. Furthermore, the hold-out validation error was also used in preliminary tests to determine training parameters, such as learning rate, number of epochs and weight decay (L2 regularization).

We note here that we conducted all training procedures and network implementations described in this paper using PyTorch, a Python library specialized in neural networks. We also used a NVidia RTX (Ray Tracing Texel eXtreme) 3080 GPU (Graphics Processing Unit), with mixed precision.

Using the segmentation masks as targets, we trained the U-Net with cross-entropy loss, stochastic gradient descent (SGD) with momentum of 0.99 and mini-batches of size 8. We began by training the network for 200 epochs with a learning rate (lr) of 10$^{-4}$. Afterwards, we changed the rate to 10$^{-5}$ and used a reduce on plateau learning rate scheduler, reducing the lr by a factor of 10 if our validation loss did not decrease in 20 epochs. We trained in this configuration for 200 epochs more.

We used mean intersection over union (IoU) to measure the U-Net test performance. IoU is a similarity measurement between two images. To calculate it we change the DNN output mask, transforming any value below 0.5 in 0 and over or equal 0.5 in 1. We then find the intersection of this binary image and the target mask (the area where both are 1), and divide it by their union (the area where the target or the output is 1). Thus, the maximum IoU is 1, when the two images are equal. Calculating the mean IoU for all testing X-rays we can quantitatively measure the DNN segmentation performance.

After this training process we achieved a mean test IoU of 0.927 in the Montgomery and Shenzen datasets. We also checked the generated images to have a qualitative measure of performance, and we found the U-Net satisfactorily segmented the lungs.

\subsubsection{Creating the masks for the segmentation dataset}
\label{masksSection}
In our segmentation dataset we only had segmentation masks for the Shenzen and Montgomery images. Thus, we still needed to create masks for the Covid-19 and pneumonia images.

We used the U-Net trained before (in the Montgomery and Shenzen datasets) to help us in this task. We began by using the DNN to generate automated masks for the pneumonia and Covid-19 images. We transformed these masks in binary, changing any value over or equal to 0.5 to 1 and below 0.5 to 0. 

Then, we manually edited the automated masks, removing imperfections and comparing them with the X-rays. The ones that were not good enough were deleted and manually redone. As in the Montgomery and Shenzen masks, we excluded areas behind the heart and used anatomical landmarks (like the ribs and the diaphragm) to create our masks. In total, we created 327 masks for the pneumonia class and 327 masks for the Covid-19 class.

\subsubsection{Training with the segmentation dataset}
With the Montgomery and Shenzen masks and the new masks for the Covid-19 and pneumonia images, we had targets for every X-ray in our segmentation dataset.

We created a new U-Net, with the same structure as the last one (\cite{unet}), to be trained using the segmentation dataset. For this process we used data augmentation (online) to avoid overfitting. All images were randomly rotated (between -40 and 40 degrees), translated (maximum of 28 pixels up or down and 28 left or right) and horizontally flipped (with a 50\% chance). This augmentation multiplied the training dataset size by 15 and we did not remove the original images. 

We trained the U-Net using cross-entropy loss, stochastic gradient descent (SGD) with momentum of 0.99 and mini-batches of size 5. We used a learning rate of 10$^{-4}$ and trained for 400 epochs (when the DNN was already overfitting). 

\begin{figure}
\includegraphics[width=0.5\textwidth]{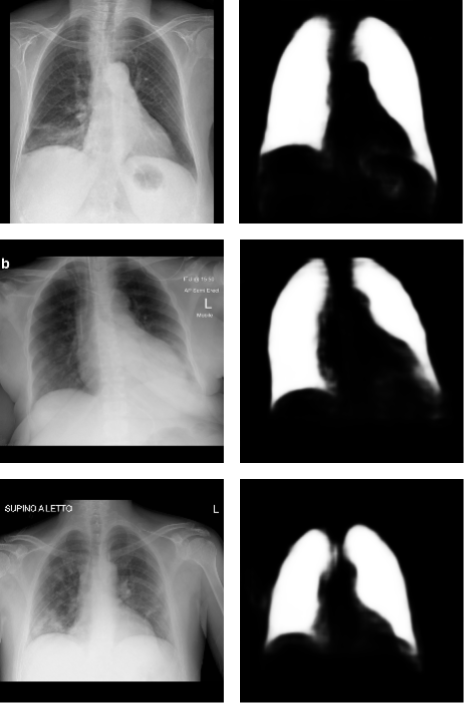}
\centering
\caption{Examples of masks (created by the U-Net) and the corresponding Covid-19 X-rays. The images were gathered from the segmentation test dataset. From top to bottom, they represent a 70-year-old female on the first day of Covid-19 symptoms, a 67-year-old female on day 8 of symptoms, and a 40-year-old male on day 10.}
\label{masksFigure}
\end{figure}

We ended up with 0.864 mean intersection over union in the test dataset. We analyzed the generated masks and found that they correctly indicated the lung areas. Most of the DNN mistakes were generating brighter regions in the gastric bubble area and in the lung region behind the heart. You can see examples of the generated masks, created with Covid-19 X-rays, in Figure \ref{masksFigure}.

\subsection{Training for classification}

We trained two DNNs for classification: a stacked network (which also performs segmentation) and a DenseNet201. The dense network and the stacked DNN classifier module have the same structure (a DenseNet201). For this reason and to better compare the networks, we trained them for classification in the same manner, described in Sections \ref{t1} and \ref{t2}.

\subsubsection{The stacked DNN creation}

To perform lung segmentation and classification we propose an architecture composed of stacked modules. The first one (segmentation module) is the U-Net, already trained on the segmentation dataset. This network receives an X-ray and outputs a segmentation mask, where high values indicate lung regions and low values refer to areas without importance. The segmentation module parameters will always be frozen when training for classification.

After the segmentation module comes the intermediate module that we designed. It applies a softmax function to the U-Net output, takes only the last dimension of the softmax result (which displays the important regions of the image with high values) and replicates it to create an image with 3 channels. Afterwards, the module performs an element-wise multiplication of this image and the input X-ray. Thus, we remove the unimportant regions from the X-ray and keep the lungs. Lastly, the module performs batch normalization on the multiplication output and our objective with this operation is to improve the DNN generalization. 

Therefore, with batch normalization the classifier input is normalized for each training mini-batch. BatchNorm is generally used to make training deep neural networks faster, by reducing the problem of internal covariance shift. However, it also makes the DNN output for a single example non-deterministic, creating a regularization effect, which improves generalization (\cite{ioffe2015batch}). Its creators discovered that the technique's regularization effect can even reduce the need for other regularization methods, like dropout (\cite{ioffe2015batch}).

The intermediate module output enters the second neural network (classification module), a DenseNet201 (\cite{DenseNet}), which predicts the chances of Covid-19, pneumonia or normal. 

We decided to use a dense neural network as our classification module because it is a large DNN with good results in image classification (\cite{DenseNet}) and because its architecture was very successful in lung disease classification, obtaining F1 scores in pneumonia detection that surpassed radiologists',  in \cite{chexnet}. Note that the F1-Score is defined as the harmonic mean between precision and recall. Considering a certain class as positive and the remaining classes as negative, we can calculate the number of true positives (tp) as the number of positive samples correctly classified, false negatives (fn) as the number of positive samples incorrectly classified as negative, and false positives (fp) as the number of negative samples incorrectly classified as positive. With these definitions in mind, it is possible to define the class precision (P), recall (R) and F1-Score. The three equations below summarize the concepts. The DenseNet201 was downloaded already pretrained on ImageNet (\cite{imagenet}), a very large image classification dataset, with millions of samples.

P=tp/(tp+fp)

R=tp/(tp+fn)

F1-Score=2 P R/(P+R)

\begin{figure}
\includegraphics[width=0.45\textwidth]{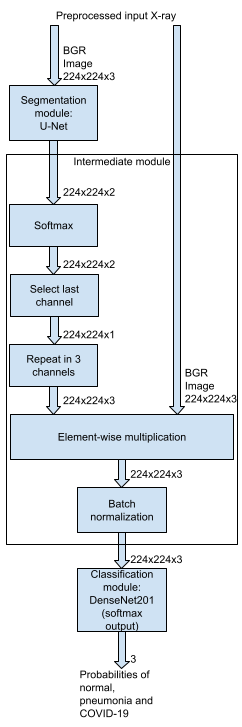}
\centering
\caption{The structure of our proposed stacked DNN, for segmentation of lungs and classification}
\label{structure}
\end{figure}

Figure \ref{structure} shows our network structure and its three modules.

\subsubsection{Pretraining with the ChestX-ray14 dataset}
\label{t1}
We trained our DNN using a twice transfer learning approach, which is similar to the one that we used in a previous Covid-19 detection study (\cite{bassi2021deep}). Another work that used twice transfer learning in a medical classification problem is \cite{DoubleMamog}, which applied the technique for mammogram classification.

Our approach consists in a transfer learning with three steps: we download ImageNet pretrained DenseNet201s (to be used as a classification DNN or the classification module of our stacked DNN), train the networks on the large ChestX-ray14 database, and then on our classification dataset (smaller, with the Covid-19, pneumonia and normal classes). We expect training on ChestX-ray14 to improve generalization of the DNNs, as it is a large X-ray database with a similar task to Covid-19 detection (classification of 14 lung diseases and healthy patients).

In the ChestX-ray14 dataset, the only augmentation technique that we applied was horizontal flipping (with 50\% chance). Unlike the augmentation we performed in the other datasets, in this case the new images substituted the original in the mini-batch (in the other datasets, the augmented images were added to the mini-batch along the original ones).

We used the test dataset reported by the database authors as our test dataset and randomly separated the remaining images, with 20\% for validation (hold-out). We did not allow two images from the same patient to be present in more than one dataset.

As classifying this dataset is a multi-label classification problem, we substituted the DNNs' last layer for one with 15 neurons and used PyTorch's binary cross-entropy loss with logits. We trained the networks using SGD, with momentum of 0.9 and mini-batches of size 64. We started by training only the last layer, with a learning rate of 10$^{-3}$, for 20 epochs. Then, we unfroze all layers (except for the segmentation module's, when training the stacked DNN) and trained for 80 epochs, with a learning rate of 10$^{-4}$. In the end of this process both DNNs were already overfitting. 

\subsubsection{Training with the classification dataset}
\label{t2}
In this step, we started with the DNNs (DenseNet201 and stacked DNN) that we trained in the ChestX-ray14 dataset and we performed the last stage of twice transfer learning: training on our classification dataset to classify the Covid-19, pneumonia and normal classes. We substituted the networks' last layer by one with 3 neurons and we added a dropout of 50\% before it (in preliminary tests, we observed that regularization improved accuracy on the external datasets).

We also utilized online data augmentation in the training dataset, to avoid overfitting and to balance the database. The augmentation process was similar to the one we used in the U-Net training (i.e., generating new images with random translation, up to 28 pixels up or down, left or right, rotation, between -40 and 40 degrees, and flipping, with 50\% chance, and not removing the original figures). In order to obtain almost the same number of images in the three classes we multiplied the number of normal and pneumonia images by 3 and of Covid-19 images by 10. We decided to use these numbers after some preliminary tests. The multiplications did not produce exactly the same number of images for each class (they created 3885 normal and pneumonia training images, and 3960 Covid-19 training images). To feed the DNN balanced mini-batches, a little amount of the images (90 of the 3960 Covid-19 augmented images and 15 of the 3885 pneumonia and normal images) were left out of training, but in every epoch a new selection of these images were made. Thus, every X-ray was used during the training process. At each epoch the neural network received 11610 training images (3870 for each class). The external validation and test datasets were not augmented.

We used cross-entropy loss, as the optimizer we chose SGD, with momentum of 0.9, and mini-batches of size 30. We trained the DNNs with hold-out validation, until we could observe a clear overfitting. We started by training only the networks' last layer, for 20 epochs, with learning rate of 10$^{-5}$ and weight decay of 0.01. We then trained all layers (except for the segmentation module, when training the stacked DNN), for 240 epochs, with weight decay of 0.05 and differential learning rates (the learning rate started at 10$^{-5}$ in the last layer was divided by 10 for each dense block before it, achieving the smallest value in the DenseNet first layer) (\cite{differentialLR}). Each epoch in this stage took about 200 s in our NVidia RTX 3080 GPU.

\subsection{Layer-wise Relevance Propagation}
DNNs are large and complex structures and it can be hard to interpret why they make decisions and classifications. Although they have a high capacity to classify images (\cite{DenseNet}), in medical applications we want to have a better understanding of how it is making its choices.

Layer-wise Relevance Propagation is a technique that makes DNNs more explainable and understandable by humans. It propagates a value called relevance from the network output layer until its first layer, creating a heatmap, with the same format as the DNN input shape. This map associates a relevance value to each each input feature (like a pixel in an image), showing how it affects the DNN output (\cite{LRP}). The relevance propagation is conservative, a neuron receives a certain amount of relevance from its posterior layer and must propagate the same quantity to the layer below it (\cite{LRPBook}). For example, if a neuron receives 10 relevance and there are three neurons in the previous layer, it can propagate relevance values of 5, 2 and 3, but not 5, 2 and 4 (as it does not sum 10). Therefore, the amount of explanation in the heatmap is directly related to what can be explained by the DNN output. We cite two uses of LRP in medical contexts: in neuroimaging (\cite{fmri-lrp}) and explaining therapy predictions (\cite{LRPMedical}). LRP has more than one rule that can be utilized to propagate relevance, and we can apply different rules in different DNN layers.

We used LRP to investigate if the DNNs were correctly interpreting symptoms of the diseases and to check if areas outside of the lungs were properly being ignored. We also think that giving these maps to clinicians along the DNN predictions may help them to evaluate the DNN classification and also provide insights about the X-rays, improving their own analysis.

We can start the relevance propagation by any output neuron and the meaning of the colors in a heatmap depends on which neuron we choose (\cite{LRPBook}). In this study, we have output neurons with indexes 0, 1 and 2, predicting the classes normal, pneumonia and Covid-19, respectively. When we start the relevance propagation by an output neuron that predicts a certain class, red areas (i.e., positive relevance) in the heatmap will show regions that the DNN associated with that class, and blue areas (i.e., negative relevance) will have been associated with the other classes. As an example, if we start LRP by the neuron that classifies the Covid-19 class (index 2), red areas in the heatmap will indicate regions associated with Covid-19, and blue areas will indicate regions associated with the other classes (normal and pneumonia). Normally we start propagation by the neuron with the highest output, i.e., the predicted class.

When analyzing the stacked DNN, we only applied LRP to the classification module, because we only wanted to know which X-ray features were important to classify the image, not to create the segmentation mask. 

To implement LRP we used the Python library iNNvestigate (\cite{innvestigate}), which already works with the DenseNet201 that we used as our classification module and as the DNN without segmentation. We chose the preset A-flat (a selection of propagation rules for the network layers), because it generated more interpretable results. To apply LRP to the classification module, we first needed to unstack our DNN. Furthermore, iNNvestigate is a library created to work with Keras and we created our DNNs using PyTorch. Thus, we used another library, called py2keras \cite{py2keras} to convert our classification module to Keras, before applying LRP. Accuracy was checked after conversion to make sure nothing went wrong.

\subsection{The Brixia Score}
\label{brixSec}

To compare our stacked DNN analysis with radiologists’, we will use the Brixia score. This scoring system, presented in \cite{brixia}, was created to grade the severity of Covid-19 cases. To score a chest X-ray, the radiologist divides the lungs into 6 parts, using two horizontal lines. The upper line is drawn at the inferior wall of the aortic line, and the other line at the level of the right pulmonary vein. If it is difficult to identify the anatomical landmarks, the authors suggest dividing the lungs into three equal zones. For each of the 6 zones, the radiologist attributes a partial score, from 0 to 3, with higher values indicating higher severity. 0 means no abnormalities in the zone, 1 means interstitial infiltrates, 2 interstitial and alveolar infiltrates, with interstitial predominance, and 3 interstitial and alveolar infiltrates, with alveolar predominance. The lines and the 6 regions are illustrated in Figure 11. Note that a single letter in red (A, B, C, D, E or F) represents a partial score for one of the 6 regions.  The overall Brixia score (from 0 to 18) is the sum of the partial scores (A+B+C+D+E+F). After the overall Brixia score, the 6 partial scores are presented, between square brackets, from A to F ([ABCDEF]). At the bottom of Figure 11, we show in red how the score is presented. Below it, in white, there is the real Brixia score for the X-ray, which presents a 72-years-old male diagnosed with Covid-19, 4 days after hospitalization. The X-ray was scored by radiologists and is presented in \cite{brixia}. The system authors discovered that the score of later deceased patients was significantly higher than from discharged patients (\cite{brixia}). 

\begin{figure}[!h]
\includegraphics[width=0.5\textwidth]{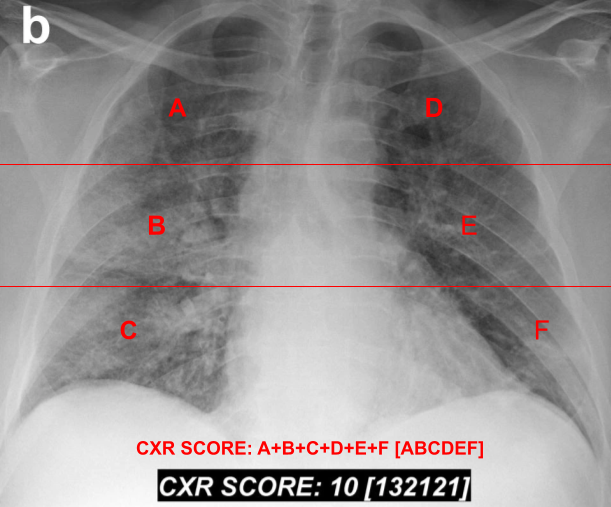}
\centering
\caption{Illustration of the lung zones for the Brixia score, and how the score is presented (bottom), based in \cite{brixia}. The Brixia score, attributed by radiologists, is presented in white. The X-ray presents a 72-year-old man with Covid-19 in the fourth day of hospitalization}
\label{brixiaFig}
\end{figure}

\subsection{The Bayesian performance evaluation}

The study in \cite{bayesianEstimator} proposed a Bayesian model to estimate the probability distribution of F1-Scores in the context of multi-class classification problems (when we have more than two classes and any sample can only be assigned to a single class).

The model can be summarized as (\cite{bayesianEstimator}):

$\bm{\upmu} \sim$ Dir$(\bm{\beta})$

$\bm{n} \sim$ Mult$(N,\bm{\upmu})$

$\bm{\theta}_{j} \sim$ Dir$(\bm{\alpha}_{j})$ \tab for j=1,...,M

$\bm{c}_{j} \sim$ Mult$(n_{j},\bm{\theta}_{j})$ \tab for j=1,...,M

$\uppsi=f(\bm{\upmu},\bm{\theta}_{1},...,\bm{\theta}_{M})$

Where N is the test dataset size (150 in this study), M the number of classes (3), Dir() represents the Dirichlet distribution and Mult() the multinomial. 

$\bm{n}$ is a random vector, with size M, $n_{j}$ estimates the number of samples in class j, if we collected a new test dataset (of size N). $\bm{\upmu}$ is also a random vector with size M and $\upmu_{j}$ indicates the probability of a new sample belonging to class j. $\bm{\beta}$ indicates the hyper-parameters of the $\bm{\upmu}$ prior distribution. Choosing $\bm{\beta}$=[1,1,1] defines a uniform prior, as we and the authors of \cite{bayesianEstimator} did.

$\bm{c}_{j}$ is a random vector of size M and $c_{j,k}$ estimates the number of class j samples classified as class k. Thus, the $c_{j,k}$ elements provide an expected confusion matrix, for a new test dataset. $\bm{\theta}_{j}$ is a random vector of size M, $\theta_{j,k}$ estimates the probability of classifying a sample from class j as class k. $\bm{\alpha}_{j}$ a vector with M hyper-parameters, defining the $\bm{\theta}_{j}$ prior distribution. As in \cite{bayesianEstimator}, we chose all elements in these vectors as 1, creating a uniform prior.

$\uppsi$ represents a function, calculated (in a deterministic manner) using the posterior probability distributions of $\bm{\upmu}$ and $\bm{\theta}$. \cite{bayesianEstimator} provides functions to estimate many performance measurements: class precision ($P_{j}$), class recall ($R_{j}$), macro-averaged F1-Score (maF1) and micro-averaged F1-Score (miF1). In a multi-class single-label classification problem, miF1 is identical to the overall accuracy (\cite{accMi}). With a balanced test dataset, like our test database, it is also identical to the average accuracy. Therefore, we used the miF1 posterior probability distribution to estimate our accuracy reliability.

We expanded the Bayesian model to also estimate the specificity for each class and their arithmetic mean. Therefore, we expressed the metrics as functions of $\bm{\upmu}$ and $\bm{\theta}$ and calculated them using these parameters posterior distributions. \cite{bayesianEstimator} defines functions for $tn_{j}$ and $fp_{j}$ (true negatives and false positives in the class j contingency table):

$tn_{j} = \sum_{u \ne j} \sum_{v \ne j} N \upmu_{u} \theta_{u,v}$

$fp_{j} = \sum_{u \ne j} N \upmu_{u} \theta_{u,j}$

Therefore, using the equations above and the definition of specificity, we can deduce the equations that define the class specificity and the mean specificity (macro-averaged)  as functions of $\bm{\upmu}$ and $\bm{\theta}$:

Specificity$_{j}=tn_{j}/(tn_{j}+fp_{j})=(\sum_{u \ne j} \sum_{v \ne j} \upmu_{u} \theta_{u,v})/(\sum_{u \ne j} \sum_{v=1}^{M} \upmu_{u} \theta_{u,v})$

Mean Specificity$ = (1/M)(\sum_{j=1}^{M}$ Specificity$_{j})$

The Bayesian model takes only the classifier confusion matrix as input, which it uses to create the likelihoods for $\bm{c}_{j}$ and $\bm{n}$.

We computed the posterior probability distributions with Markov chain Monte Carlo (MCMC), utilizing the Python library PyMC3 (\cite{pymc3}). We used the No-U-Turn Sampler (\cite{NUTS}), with 4 chains, 10000 tuning samples and 100000 samples after tuning.

\section{Results}
Table \ref{tableStacked} shows the confusion matrix for our stacked DNN, and table \ref{No Stack confusion} for the DNN without segmentation (we created both matrices using the external test database). 

Tables \ref{tabStack} and \ref{tabNoStack} show performance metrics in the external test dataset, for the DNNs with and without segmentation, respectively. In the second column (Score) we show performance scores, calculated in the traditional and deterministic manner, using the confusion matrix. The other columns refer to statistics of the metrics' posterior distributions, estimated using Bayesian inference. They are: mean, standard deviation (std), Monte Carlo error, and 95\% highest density interval (HDI). The HDI is defined as an interval with 95\% of the distribution probability mass and any point in this interval has a probability that is higher than any point outside the HDI.

\begin{table}[]
\centering
\begin{tabular}{ll|l|l|l|}
\cline{3-5}
                                                                                            &           & \multicolumn{3}{l|}{Predicted Class} \\ \cline{3-5} 
                                                                                            &           & Normal    & Pneumonia   & Covid-19   \\ \hline
\multicolumn{1}{|l|}{\multirow{3}{*}{\begin{tabular}[c]{@{}l@{}}Real\\ Class\end{tabular}}} & Normal    & 38        & 7           & 5          \\ \cline{2-5} 
\multicolumn{1}{|l|}{}                                                                      & Pneumonia & 8         & 32          & 10         \\ \cline{2-5} 
\multicolumn{1}{|l|}{}                                                                      & Covid-19  & 2         & 0           & 48         \\ \hline
\end{tabular}
\caption{Stacked DNN confusion matrix}
\label{tableStacked}
\end{table}

\begin{table}[]
\centering
\begin{tabular}{ll|l|l|l|}
\cline{3-5}
                                                                                            &           & \multicolumn{3}{l|}{Predicted Class} \\ \cline{3-5} 
                                                                                            &           & Normal    & Pneumonia   & Covid-19   \\ \hline
\multicolumn{1}{|l|}{\multirow{3}{*}{\begin{tabular}[c]{@{}l@{}}Real\\ Class\end{tabular}}} & Normal    & 43        & 0           & 7          \\ \cline{2-5} 
\multicolumn{1}{|l|}{}                                                                      & Pneumonia & 14        & 24          & 12         \\ \cline{2-5} 
\multicolumn{1}{|l|}{}                                                                      & Covid-19  & 6         & 0           & 44         \\ \hline
\end{tabular} 
\caption{Confusion matrix for the DNN without segmentation }
\label{No Stack confusion}
\end{table}

\begin{table}[]
\centering
\begin{tabular}{|l|l|l|l|l|l|}
\hline
Metric                     & Score & Mean  & std   & MC error & 95\% HDI          \\ \hline
Mean Accuracy or miF1      & 0.787 & 0.761 & 0.034 & 0.0      & {[}0.695,0.826{]} \\ \hline
Macro-averaged F1-Score    & 0.781 & 0.754 & 0.034 & 0.0      & {[}0.687,0.82{]}  \\ \hline
Macro-averaged Precision   & 0.791 & 0.764 & 0.034 & 0.0      & {[}0.697,0.829{]} \\ \hline
Macro-averaged Recall      & 0.787 & 0.761 & 0.032 & 0.0      & {[}0.698,0.823{]} \\ \hline
Macro-averaged Specificity & 0.893 & 0.88  & 0.017 & 0.0      & {[}0.848,0.912{]} \\ \hline
Normal Precision           & 0.792 & 0.765 & 0.059 & 0.0      & {[}0.648,0.877{]} \\ \hline
Normal Recall              & 0.76  & 0.736 & 0.06  & 0.0      & {[}0.617,0.85{]}  \\ \hline
Normal F1-Score            & 0.776 & 0.748 & 0.048 & 0.0      & {[}0.654,0.839{]} \\ \hline
Normal Specificity         & 0.9   & 0.887 & 0.031 & 0.0      & {[}0.825,0.943{]} \\ \hline
Pneumonia Precision        & 0.821 & 0.786 & 0.063 & 0.0      & {[}0.66,0.902{]}  \\ \hline
Pneumonia Recall           & 0.64  & 0.623 & 0.066 & 0.0      & {[}0.493,0.75{]}  \\ \hline
Pneumonia F1-Score         & 0.719 & 0.692 & 0.054 & 0.0      & {[}0.586,0.795{]} \\ \hline
Pneumonia Specificity      & 0.93  & 0.915 & 0.027 & 0.0      & {[}0.861,0.964{]} \\ \hline
Covid-19 Precision         & 0.762 & 0.742 & 0.054 & 0.0      & {[}0.636,0.844{]} \\ \hline
Covid-19 Recall            & 0.96  & 0.925 & 0.036 & 0.0      & {[}0.854,0.985{]} \\ \hline
Covid-19 F1-Score          & 0.85  & 0.822 & 0.038 & 0.0      & {[}0.746,0.894{]} \\ \hline
Covid-19 Specificity       & 0.85  & 0.84  & 0.035 & 0.0      & {[}0.769,0.906{]} \\ \hline
\end{tabular}
\caption{Performance metrics for the DNN with segmentation. The score values are traditional point estimates. The other values were obtained with Bayesian inference. }
\label{tabStack}
\end{table}

\begin{table}[]
\centering
\begin{tabular}{|l|l|l|l|l|l|}
\hline
Metric                     & Score & Mean  & std   & MC error & 95\% HDI          \\ \hline
Mean Accuracy or miF1      & 0.74  & 0.717 & 0.036 & 0.0      & {[}0.646,0.786{]} \\ \hline
Macro-averaged F1-Score    & 0.729 & 0.705 & 0.037 & 0.0      & {[}0.632,0.776{]} \\ \hline
Macro-averaged Precision   & 0.794 & 0.758 & 0.032 & 0.0      & {[}0.696,0.82{]}  \\ \hline
Macro-averaged Recall      & 0.74  & 0.717 & 0.033 & 0.0      & {[}0.653,0.781{]} \\ \hline
Macro-averaged Specificity & 0.87  & 0.858 & 0.017 & 0.0      & {[}0.825,0.891{]} \\ \hline
Normal Precision           & 0.683 & 0.667 & 0.058 & 0.0      & {[}0.553,0.779{]} \\ \hline
Normal Recall              & 0.86  & 0.83  & 0.051 & 0.0      & {[}0.728,0.924{]} \\ \hline
Normal F1-Score            & 0.761 & 0.738 & 0.045 & 0.0      & {[}0.647,0.824{]} \\ \hline
Normal Specificity         & 0.8   & 0.792 & 0.039 & 0.0      & {[}0.714,0.867{]} \\ \hline
Pneumonia Precision        & 1.0   & 0.926 & 0.05  & 0.0      & {[}0.829,0.998{]} \\ \hline
Pneumonia Recall           & 0.48  & 0.472 & 0.068 & 0.0      & {[}0.34,0.605{]}  \\ \hline
Pneumonia F1-Score         & 0.649 & 0.622 & 0.063 & 0.0      & {[}0.497,0.743{]} \\ \hline
Pneumonia Specificity      & 1.0   & 0.981 & 0.013 & 0.0      & {[}0.955,1.0{]}   \\ \hline
Covid-19 Precision         & 0.698 & 0.682 & 0.057 & 0.0      & {[}0.569,0.792{]} \\ \hline
Covid-19 Recall            & 0.88  & 0.849 & 0.049 & 0.0      & {[}0.752,0.938{]} \\ \hline
Covid-19 F1-Score          & 0.779 & 0.755 & 0.044 & 0.0      & {[}0.667,0.839{]} \\ \hline
Covid-19 Specificity       & 0.81  & 0.802 & 0.039 & 0.0      & {[}0.726,0.876{]} \\ \hline
\end{tabular}
\caption{Performance  metrics  for  the  DNN  without  segmentation. The  score  values  are traditional point estimates.  The other values were obtained with Bayesian inference.}
\label{tabNoStack}
\end{table}

We calculated, with the test dataset, the multi-class area under the ROC curve (AUC) using macro averaging and the pairwise comparisons approach from \cite{MulticlassAUC}. The stacked DNN achieved 0.917 AUC and the DenseNet201, 0.906. We do not present interval estimations for multi-class AUC because defining its confidence interval is not a simple task, and bootstrapping is the suggested method for it (\cite{MulticlassAUC}). We can not use bootstrapping in this study, as we are using an external test dataset and we have a small amount of Covid-19 X-rays.

\begin{figure}[h]
\includegraphics[width=1\textwidth]{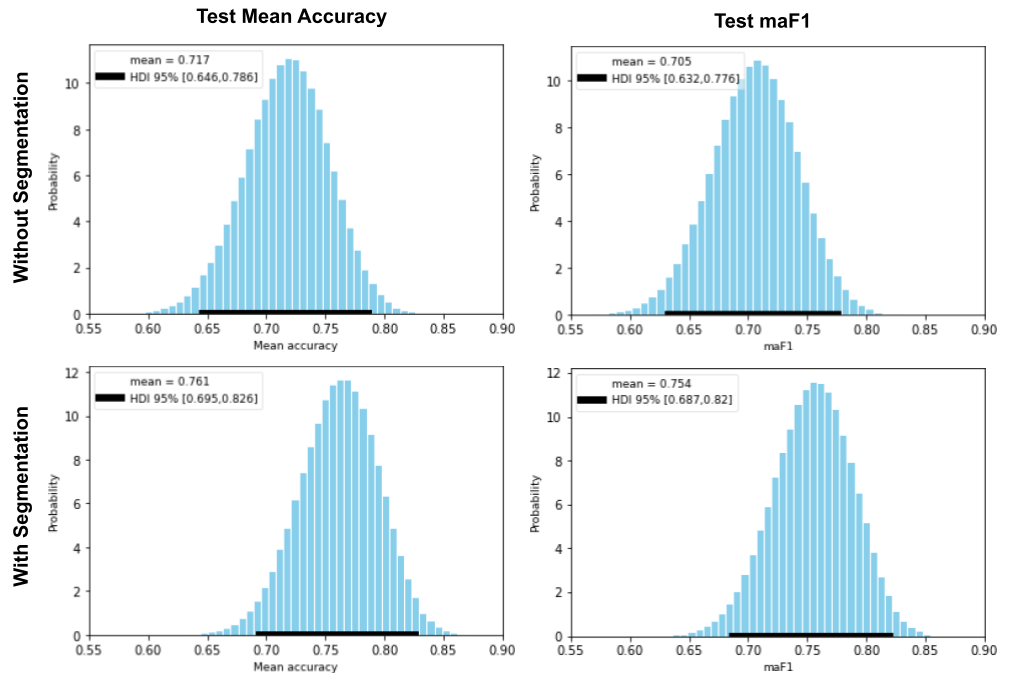}
\centering
\caption{Posterior probability density estimations for mean accuracy (left) and macro-averaged F1-Score (right), considering the DNNs with (bottom) and without segmentation (top)}
\label{estimations}
\end{figure}

\begin{figure}[!h]
\includegraphics[width=1\textwidth]{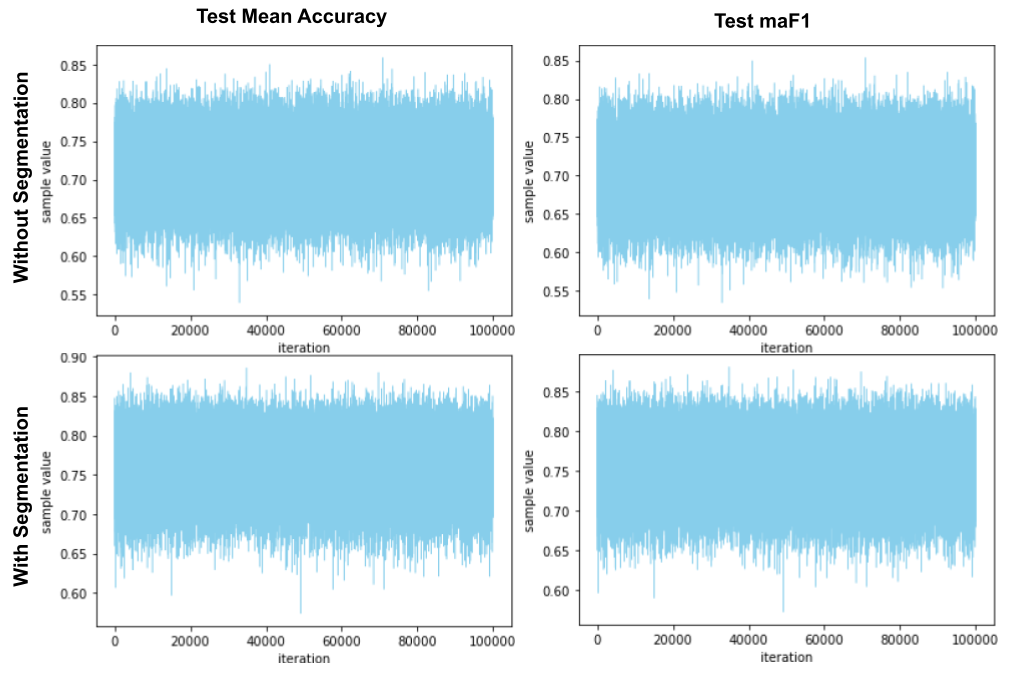}
\centering
\caption{Trace plots for mean accuracy (left) and macro-averaged F1-Score (right), considering the DNNs with (bottom) and without segmentation (top)}
\label{traces}
\end{figure}

In Figure \ref{estimations} we show the Bayesian estimations of mean accuracy (equal to miF1) and macro-averaged F1-Score. In Figure \ref{traces} we display the corresponding trace plots (for only one MCMC chain). These plots exclude the tuning samples.

Our trained DNNs are available for download at https://github.com/PedroRASB/Covid-19-detection-with-lung-segmentation.

\subsection{Discussion}

In a previous study, we utilized a dataset that was very similar to our classification training database. We also trained dense neural networks (without segmentation), but we did not perform validation and testing on an external dataset (\cite{bassi2021deep}). There, we could achieve accuracies above 90\%, as is common in many Covid-19 detection studies, which also use internal validation, i.e., they randomly divide a single dataset in testing, validation and training (\cite{reviewCovid}). Furthermore, in preliminary tests using the stacked DNN that we proposed here, but without external validation, we could also achieve accuracies above 90\%. We note that, in our previous study and in the preliminary tests, our classification training database was divided in three datasets (for training, validation and test) and two images from the same patient were not allowed to be present in two different datasets. We conclude that evaluating DNNs in an external dataset can show significantly smaller accuracies, indicating that bias can hinder generalization when working with mixed datasets, and internal validation results may not reflect performance when analyzing data from other hospitals and locations.

Furthermore, when we compare the results of our stacked DNN and the DenseNet201, we observe that segmentation has an effect on the model generalization capability, increasing mean accuracy score on the external test dataset by 4.7\%, and the Bayesian estimation mean by 4.4\%.

Some works have also used lung segmentation for Covid-19 detection in chest X-rays. A recent study (\cite{otherSegment}) used a modified U-Net to segment the X-rays beforehand, it then applied an image enhancement technique (like histogram equalization) and classified the segmented X-ray with different DNNs. Like in this study, their work utilized a mixed database, but, unlike our work, they constructed their test dataset randomly, using 5-fold cross validation. As can be seen in other works that applied internal validation (\cite{reviewCovid}), their study obtained high accuracies, around 95\%. But, surprisingly, their results showed that lung segmentation reduced test accuracy and F1-scores (by about 1\%). This result strongly contrasts with our findings (4.7\% accuracy increment with segmentation), and, although the utilization of our intermediate module might have positively influenced our performances with segmentation, we do not think that it is the main cause for this discrepancy. Instead, we think that the different test methodologies in the two papers caused the discrepant results. In our study, lung segmentation reduced dataset bias, improving generalization and the results on the external test dataset. However, this reduction of dataset bias may actually decrease performances when they are measured with internal validation, possibly explaining why lung segmentation reduced accuracy and F1-Score in \cite{otherSegment}.

The normal class specificity shows the percentage of unhealthy patients that were not classified as healthy. The score value of 90\%, in table \ref{tabStack}, indicates that a relatively low number of the patients with a disease were miss-classified as healthy by our model.

We note that the 95\% HDIs are relatively large, e.g., for mean accuracy with the stacked DNN the interval length is 0.131. This can also be observed in Figure \ref{estimations}.  We suppose that the strongest reason for the large intervals is the small size of the test dataset, as using more test samples would increase the performance metrics confidence. 

\subsection{LRP and comparison with radiologists' analysis (using the Brixia score)}

We propose comparing X-rays scored by radiologists, using the Brixia score, with heatmaps, created by LRP. The maps show how much relevance in classification each part of the X-rays has. Therefore, if we start the propagation by the neuron that classifies Covid-19, areas that have larger and darker red regions indicate where the DNN found more Covid-19 symptoms. Checking these areas' partial Brixia scores may indicate if the DNNs look for the same signs of Covid-19 as radiologists do. Furthermore, more severe cases of Covid-19 may show stronger symptoms and could increase the Covid-19 probability predicted by the DNN. Therefore, we may also be able to check if there is a correlation between images with high Brixia scores and the higher predicted probabilities.

\begin{figure}
\includegraphics[width=1\textwidth]{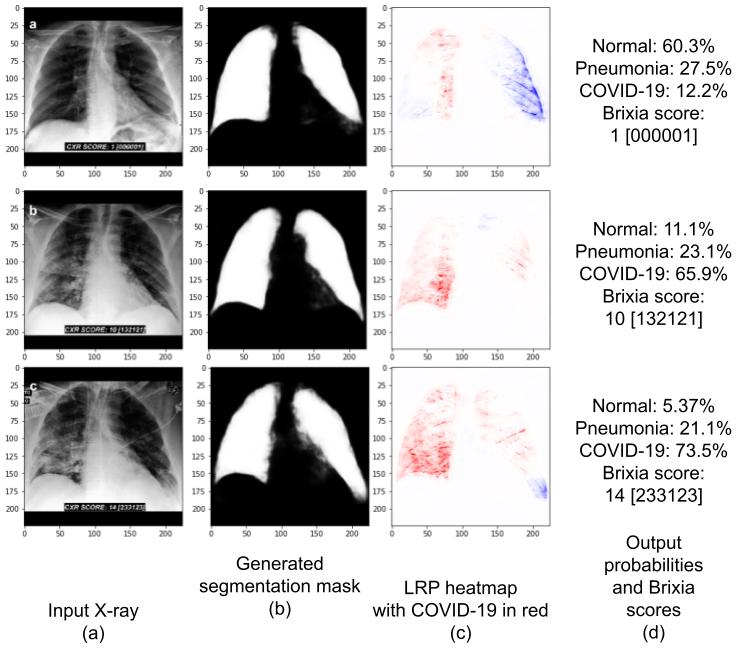}
\centering
\caption{X-rays (column a) are from a male 72-year-old Covid-19 patient. From top to bottom, the X-rays were created in the first, fourth and fifth days of hospitalization. Masks (column b), heatmaps (column c), and class probabilities (column d) consider the stacked DNN. In the heatmaps, red colors indicate areas that the DNN associated to Covid-19, while blue areas were associated to the pneumonia or normal classes. We observe a clear correlation between Brixia scores values and red regions in the heatmaps. Moreover, as the disease progressed, Brixia scores increased, the DNN predicted higher Covid-19 probability, and heatmaps became more red}
\label{relevance1}
\end{figure}

Besides presenting the scoring system, \cite{brixia} also shows examples of Covid-19 X-rays, already scored by radiologists. These images are also part of our training dataset. In Figure \ref{relevance1} we analyze, with our stacked DNN, three of them (the ones that had nothing written over the lungs). The X-rays displayed in this section were processed as indicated in Methods Section 5. The Figure presents the X-rays, the generated segmentation masks, the LRP heatmaps, the network outputs and the Brixia scores (given by radiologists), with the partial scores in brackets. We note that relevance propagation began at the neuron that classifies Covid-19, therefore, red areas indicate regions that the DNN associated with Covid-19, while blue areas were associated with the pneumonia or the normal class. 

All X-rays in Figure \ref{relevance1} were taken from a 72-years-old man diagnosed with Covid-19. The one in the first row is from the day of admission, one day after the onset of fever (\cite{brixia}). We observe that the X-ray shows little signs of Covid-19, as the Brixia score is very low, at one. This should make classification more challenging, and, indeed, our DNN could not correctly classify this image, predicting the normal class, but with only 60.3\% probability. The patient had a rapid disease progression, the second and third rows show X-rays at days 4 and 5 post-hospitalization, respectively (\cite{brixia}). Our DNN correctly classified both X-rays, with Covid-19 probabilities of 65.9\% and 73.5\%. 

Unlike the Brixia score, our network is not designed to analyze disease severity. But we observe that X-rays showing more severe and apparent symptoms (thus, with higher Brixia scores) also increase the DNN confidence for the Covid-19 class. In Figure \ref{relevance1} we see that the higher the Brixia score, the higher the Covid-19 predicted probability. This indicates a similarity between the symptoms that the radiologists look for and the ones that our DNN analyses. 

An analysis of the partial scores and the heatmaps of the two correctly classified X-rays in Figure \ref{relevance1} also corroborates with the conclusion above. In both heatmaps, we observe more relevance in the right lung, and it also has higher Brixia scores. The middle heatmap shows that, in the right lung, the DNN found more Covid-19 signs in regions B and C, which also have higher partial Brixia scores; in the left lung, we see more relevance in the E region, the one with the higher partial score. In the lower heatmap, in the right lung, there is again more relevance in regions B and C, which also present higher Brixia scores. The F region of the lower heatmap in Figure \ref{relevance1} has 3 Brixia score, but is blue in our heatmap. The reason for this is that the region was mostly associated, by our DNN, with the pneumonia class (this region is very red if we start LRP by the neuron that classifies pneumonia). 

\begin{figure}
\includegraphics[width=1\textwidth]{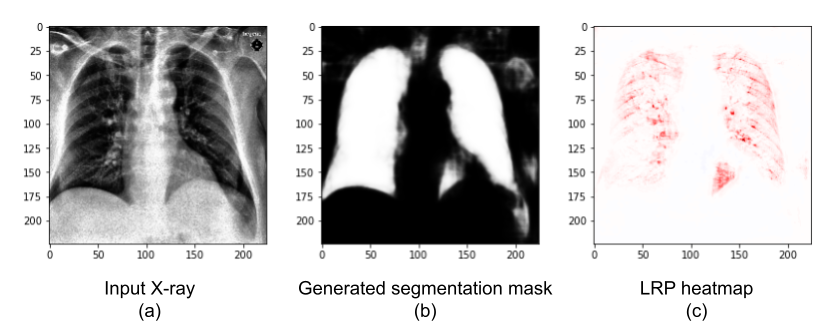}
\centering
\caption{The X-ray (a) is an image from our external test dataset (unlike Figure \ref{relevance1}, which presented training X-rays), correctly classified as Covid-19. It presents a male patient in the first day of Covid-19 symptoms. The mask (b) and the heatmap (c) were created with the stacked DNN. Red colors indicate areas that the DNN associated to Covid-19, while blue areas were associated to the classes pneumonia or normal}
\label{relevance2}
\end{figure}

LRP analysis also showed that our segmentation module and intermediate module work as intended, maintaining almost all relevance in the lung regions (as can be seen in Figures \ref{relevance1} and \ref{relevance2}). Figure \ref{relevance2} also analyzes the stacked DNN. It shows a Covid-19 input X-ray, the generated mask and LRP heatmap. But, unlike in Figure \ref{relevance1}, this radiography is from the external test dataset. We observe that the segmentation mask is not perfect, but the areas outside the lungs are not very bright and are mostly ignored by the DNN, as the heatmap shows. Again, this X-ray was correctly classified (89.6\% probability of Covid-19) and the red areas in the heatmap were associated, by the neural network, with the Covid-19 class.

\begin{figure}
\includegraphics[width=0.75\textwidth]{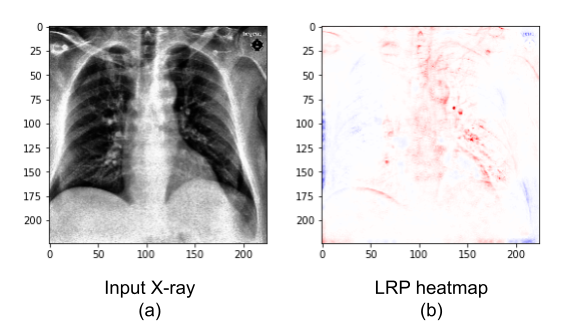}
\centering
\caption{Covid-19 X-ray (a) and heatmap (b). Unlike Figures \ref{relevance1} and \ref{relevance2}, this heatmap was created with the DNN without lung segmentation. The X-ray is an image from our external test dataset, correctly classified by the network. It presents a male patient in the first day of Covid-19 symptoms. Red colors indicate areas that the DNN associated to the Covid-19 class, while blue areas were associated to pneumonia or normal}
\label{relevance3}
\end{figure}

We can further understand the differences between the two DNNs (with and without segmentation) when we analyze them using Layer-wise Relevance Propagation. Therefore, we show, in Figure \ref{relevance3}, a LRP analysis for the same X-ray in Figure \ref{relevance2}, but created using the DenseNet201 (without segmentation) instead of the stacked DNN. We note that this DNN correctly classifies the image, but it assigned a much lower Covid-19 probability, of 46.2\%. Red areas in the map were associated with the Covid-19 class, while blue areas were associated with the other classes.

We observe, in Figure \ref{relevance3}, that there is relevance outside of the lungs. Its existence may explain why the stacked DNN has better generalization (4.7\% higher accuracy on the external test dataset) than the network without segmentation. The relevance outside of the important areas might indicate dataset biases learned by the DNN. However, some Covid-19 signs, indicated in the heatmap in Figure \ref{relevance2} can still be seen on Figure \ref{relevance3} (mostly on the left lung).

\section{Conclusion}

First, we observe that our mean accuracy score, on the external test dataset, using the stacked DNN was 78.7\% and, without segmentation, 74\%. These values are significantly lower than the accuracies calculated using internal validation (i.e., randomly separating a database in test, validation and training datasets). Our previous study (which used a database similar to our current classification training dataset and utilized internal validation, without lung segmentation) (\cite{bassi2021deep}), and many other works that detected Covid-19 using DNNs without external validation (\cite{reviewCovid}) showed accuracies above 90\%. This performance discrepancy may indicate that utilizing mixed datasets creates bias, which improves internal validation accuracies and performance metrics, as the study in \cite{ArtigoCritico} suggests. These extremely high accuracies may not hold up when images from other hospitals, locations and datasets are analyzed, as we have seen in this work.

The utilization of segmentation, performed by our stacked DNN architecture, improved generalization, increasing mean accuracy score on the external test dataset by 4.7\% (or 4.4\%, when considering the Bayesian estimations means). Other techniques that may have helped mitigating mixed dataset bias in this study were: histogram equalization (in the input X-rays), batch normalization (in our intermediate module), removing pediatric patients from the datasets (because the youngest patient in the Covid-19 class is 20 years old), utilizing an external validation dataset, regularization (dropout and weight decay), twice transfer learning and data augmentation.

Bayesian estimation of the DNNs' performance metrics allowed us to quantify the reliability of the metrics. We observed relatively large 95\% highest density intervals, caused by the small size of the test dataset (150 images). This emphasizes both the importance of making interval estimations in the context of Covid-19 detection, and how beneficial would larger Covid-19 X-ray databases be.

Layer-wise Relevance Propagation allowed us to generate heatmaps and analyze how our DNNs performed their classification. The stacked DNN heatmaps indicated that the networks successfully ignored areas outside the lungs, because these regions' relevance was very small (showing almost no color in the maps). Comparing X-rays scored by radiologists using the Brixia score to our stacked DNN outputs and heatmaps showed that, normally, regions with higher partial scores also had higher Covid-19 LRP relevance. Also, X-rays with higher overall scores were associated with higher Covid-19 predicted probabilities. This may indicate that radiologists and our stacked DNN look for the same signs of Covid-19 in a radiography. Unfortunately, our DNN could not correctly classify an X-ray where radiologists also found few symptoms of Covid-19 (the upper X-ray in Figure \ref{relevance1}, with a small overall Brixia score, of only 1). 

Performing LRP in the DenseNet201 without segmentation indicated that, although lung areas were relevant and taken into account, the DNN also paid attention to regions outside of the lungs. This again suggests that segmentation can reduce dataset bias and improve generalization.

Although we conclude that mixed dataset bias is significant, our DNNs' performance on an external dataset and LRP analysis indicate that it can be partially avoided. On the external test dataset our stacked network had 0.916 AUC and, using the Bayesian model, we estimated a macro-averaged F1-Score with mean of 0.754 and 95\% highest density interval of [0.687,0.82]. 

This study shows the need for a large, open and high quality Covid-19 X-ray database, with all classes collected from the same sources, to better avoid dataset bias, improve generalization and increase performance metrics reliability. Our DNNs' performance in the external dataset suggests that, even with small and mixed datasets, DNNs can be successfully trained to detect Covid-19, if appropriate measures to avoid bias are taken. However, we must note that even though we utilized an external test dataset, clinical tests are needed to further ensure that the performances we observed in this study are replicable in a real-world scenario.

This work employed a non-standard testing strategy, evaluating the DNNs on an external, out-of-distribution dataset. Therefore, it mitigated the effect of bias in the reported results, and more realistically assessed the potential of deep learning to become an auxiliary tool to help clinicians in Covid-19 detection. Moreover, using this evaluation strategy we demonstrated the importance of lung segmentation to improve DNN generalization, a capability that is paramount for the neural network applicability in a real clinical scenario. Finally, our novel analysis with the Brixia score and LRP heatmaps allowed a more profound understanding of the deep neural network decision rules, increasing its trustworthiness, a quality that is crucial for medical applications. 

\section{Acknowledgments}
This work was partially supported by Conselho Nacional de Desenvolvimento Científico e Tecnológico (process 308811/2019-4) and Coordenação de Aperfeiçoamento de Pessoal de Nível Superior.

\section{Statements and Declarations}
The authors declare no competing interests.

\bibliography{mybibfile}

\begin{thebibliography}{36}
\providecommand{\natexlab}[1]{#1}
\providecommand{\url}[1]{\texttt{#1}}
\expandafter\ifx\csname urlstyle\endcsname\relax
  \providecommand{\doi}[1]{doi: #1}\else
  \providecommand{\doi}{doi: \begingroup \urlstyle{rm}\Url}\fi

\bibitem[Alber et~al.(2019)Alber, Lapuschkin, Seegerer, H{\"a}gele, Sch{\"u}tt,
  Montavon, Samek, M{\"u}ller, D{\"a}hne, and Kindermans]{innvestigate}
M.~Alber, S.~Lapuschkin, P.~Seegerer, M.~H{\"a}gele, K.~T. Sch{\"u}tt,
  G.~Montavon, W.~Samek, K.-R. M{\"u}ller, S.~D{\"a}hne, and P.-J. Kindermans.
\newblock innvestigate neural networks!
\newblock \emph{Journal of Machine Learning Research}, 20\penalty0
  (93):\penalty0 1--8, 2019.
\newblock URL \url{http://jmlr.org/papers/v20/18-540.html}.

\bibitem[Bach et~al.(2015)Bach, Binder, Montavon, Klauschen, Müller, and
  Samek]{LRP}
S.~Bach, A.~Binder, G.~Montavon, F.~Klauschen, K.-R. Müller, and W.~Samek.
\newblock On pixel-wise explanations for non-linear classifier decisions by
  layer-wise relevance propagation.
\newblock \emph{PLOS ONE}, 10\penalty0 (7):\penalty0 1--46, 07 2015.
\newblock \doi{10.1371/journal.pone.0130140}.
\newblock URL \url{https://doi.org/10.1371/journal.pone.0130140}.

\bibitem[Bassi and Attux(2021)]{bassi2021deep}
P.~R. A.~S. Bassi and R.~Attux.
\newblock A deep convolutional neural network for covid-19 detection using
  chest x-rays.
\newblock \emph{Research on Biomedical Engineering}, Apr 2021.
\newblock ISSN 2446-4740.
\newblock \doi{10.1007/s42600-021-00132-9}.
\newblock URL \url{http://dx.doi.org/10.1007/s42600-021-00132-9}.

\bibitem[Borghesi and Maroldi(2020)]{brixia}
A.~Borghesi and R.~Maroldi.
\newblock Covid-19 outbreak in italy: experimental chest x-ray scoring system
  for quantifying and monitoring disease progression.
\newblock \emph{La radiologia medica}, 125, 05 2020.
\newblock \doi{10.1007/s11547-020-01200-3}.

\bibitem[{Cai} et~al.(2018){Cai}, {Liu}, and {Guo}]{DoubleMamog}
Q.~{Cai}, X.~{Liu}, and Z.~{Guo}.
\newblock Identifying architectural distortion in mammogram images via a
  se-densenet model and twice transfer learning.
\newblock In \emph{2018 11th International Congress on Image and Signal
  Processing, BioMedical Engineering and Informatics (CISP-BMEI)}, pages 1--6,
  2018.
\newblock \doi{10.1109/CISP-BMEI.2018.8633197}.
\newblock URL \url{https://doi.org/10.1109/CISP-BMEI.2018.8633197}.

\bibitem[Candemir et~al.(2014)Candemir, Jaeger, Palaniappan, Musco, Singh, Xue,
  Karargyris, Antani, Thoma, and Mcdonald]{ChineseDataset1}
S.~Candemir, S.~Jaeger, K.~Palaniappan, J.~Musco, R.~Singh, Z.~Xue,
  A.~Karargyris, S.~Antani, G.~Thoma, and C.~Mcdonald.
\newblock Lung segmentation in chest radiographs using anatomical atlases with
  nonrigid registration.
\newblock \emph{IEEE Transactions on Medical Imaging}, 33:\penalty0 577--590,
  02 2014.
\newblock \doi{10.1109/TMI.2013.2290491}.

\bibitem[Cohen et~al.(2020)Cohen, Morrison, and Dao]{GitCovidSet}
J.~P. Cohen, P.~Morrison, and L.~Dao.
\newblock Covid-19 image data collection.
\newblock \emph{arXiv 2003.11597}, 2020.
\newblock URL \url{https://github.com/ieee8023/covid-chestxray-dataset}.

\bibitem[{Deng} et~al.(2009){Deng}, {Dong}, {Socher}, {Li}, {Kai Li}, and {Li
  Fei-Fei}]{imagenet}
J.~{Deng}, W.~{Dong}, R.~{Socher}, L.~{Li}, {Kai Li}, and {Li Fei-Fei}.
\newblock Imagenet: A large-scale hierarchical image database.
\newblock In \emph{2009 IEEE Conference on Computer Vision and Pattern
  Recognition}, pages 248--255, 2009.
\newblock \doi{10.1109/CVPR.2009.5206848}.
\newblock URL \url{https://doi.org/10.1109/CVPR.2009.5206848}.

\bibitem[Guan et~al.(2020)Guan, Ni, Hu, Liang, Ou, He, Liu, Shan, Lei, Hui, Du,
  Li, Zeng, Yuen, Chen, Tang, Wang, Chen, Xiang, Li, Wang, Liang, Peng, Wei,
  Liu, Hu, Peng, Wang, Liu, Chen, Li, Zheng, Qiu, Luo, Ye, Zhu, and
  Zhong]{clinicalCOVID}
W.-j. Guan, Z.-y. Ni, Y.~Hu, W.-h. Liang, C.-q. Ou, J.-x. He, L.~Liu, H.~Shan,
  C.-l. Lei, D.~S. Hui, B.~Du, L.-j. Li, G.~Zeng, K.-Y. Yuen, R.-c. Chen, C.-l.
  Tang, T.~Wang, P.-y. Chen, J.~Xiang, S.-y. Li, J.-l. Wang, Z.-j. Liang, Y.-x.
  Peng, L.~Wei, Y.~Liu, Y.-h. Hu, P.~Peng, J.-m. Wang, J.-y. Liu, Z.~Chen,
  G.~Li, Z.-j. Zheng, S.-q. Qiu, J.~Luo, C.-j. Ye, S.-y. Zhu, and N.-s. Zhong.
\newblock Clinical characteristics of coronavirus disease 2019 in china.
\newblock \emph{New England Journal of Medicine}, 382\penalty0 (18):\penalty0
  1708--1720, 2020.
\newblock \doi{10.1056/NEJMoa2002032}.
\newblock URL \url{https://doi.org/10.1056/NEJMoa2002032}.

\bibitem[Hand and Till(2001)]{MulticlassAUC}
D.~J. Hand and R.~J. Till.
\newblock A simple generalisation of the area under the roc curve for multiple
  class classification problems.
\newblock \emph{Mach. Learn.}, 45\penalty0 (2):\penalty0 171–186, Oct. 2001.
\newblock ISSN 0885-6125.
\newblock \doi{10.1023/A:1010920819831}.
\newblock URL \url{https://doi.org/10.1023/A:1010920819831}.

\bibitem[Heo et~al.(2019)Heo, Kim, Yun, Lim, Kim, Nam, Park, Jung, and
  Yoon]{TuberculosisUNet}
S.-J. Heo, Y.~Kim, S.~Yun, S.-S. Lim, J.~Kim, C.-M. Nam, E.-C. Park, I.~Jung,
  and J.-H. Yoon.
\newblock Deep learning algorithms with demographic information help to detect
  tuberculosis in chest radiographs in annual workers’ health examination
  data.
\newblock \emph{International Journal of Environmental Research and Public
  Health}, 16:\penalty0 250, 01 2019.
\newblock \doi{10.3390/ijerph16020250}.
\newblock URL \url{https://doi.org/10.3390/ijerph16020250}.

\bibitem[Homan and Gelman(2014)]{NUTS}
M.~D. Homan and A.~Gelman.
\newblock The no-u-turn sampler: Adaptively setting path lengths in hamiltonian
  monte carlo.
\newblock \emph{J. Mach. Learn. Res.}, 15\penalty0 (1):\penalty0 1593–1623,
  Jan. 2014.
\newblock ISSN 1532-4435.

\bibitem[Howard and Ruder(2018)]{differentialLR}
J.~Howard and S.~Ruder.
\newblock Universal language model fine-tuning for text classification.
\newblock pages 328--339, 01 2018.
\newblock \doi{10.18653/v1/P18-1031}.
\newblock URL \url{https://doi.org/10.18653/v1/P18-1031}.

\bibitem[{Huang} et~al.(2017){Huang}, {Liu}, {Van Der Maaten}, and
  {Weinberger}]{DenseNet}
G.~{Huang}, Z.~{Liu}, L.~{Van Der Maaten}, and K.~Q. {Weinberger}.
\newblock Densely connected convolutional networks.
\newblock In \emph{2017 IEEE Conference on Computer Vision and Pattern
  Recognition (CVPR)}, pages 2261--2269, 2017.
\newblock \doi{10.1109/CVPR.2017.243}.
\newblock URL \url{https://doi.org/10.1109/CVPR.2017.243}.

\bibitem[Ioffe and Szegedy(2015)]{ioffe2015batch}
S.~Ioffe and C.~Szegedy.
\newblock Batch normalization: Accelerating deep network training by reducing
  internal covariate shift, 2015.

\bibitem[Irvin et~al.(2019)Irvin, Rajpurkar, Ko, Yu, Ciurea-Ilcus, Chute,
  Marklund, Haghgoo, Ball, Shpanskaya, Seekins, Mong, Halabi, Sandberg, Jones,
  Larson, Langlotz, Patel, Lungren, and Ng]{irvin2019chexpert}
J.~Irvin, P.~Rajpurkar, M.~Ko, Y.~Yu, S.~Ciurea-Ilcus, C.~Chute, H.~Marklund,
  B.~Haghgoo, R.~Ball, K.~Shpanskaya, J.~Seekins, D.~Mong, S.~Halabi,
  J.~Sandberg, R.~Jones, D.~Larson, C.~Langlotz, B.~Patel, M.~Lungren, and
  A.~Ng.
\newblock Chexpert: A large chest radiograph dataset with uncertainty labels
  and expert comparison.
\newblock \emph{Proceedings of the AAAI Conference on Artificial Intelligence},
  33:\penalty0 590--597, 07 2019.
\newblock \doi{10.1609/aaai.v33i01.3301590}.
\newblock URL \url{https://doi.org/10.1609/aaai.v33i01.3301590}.

\bibitem[{Jaeger} et~al.(2014){Jaeger}, {Karargyris}, {Candemir}, {Folio},
  {Siegelman}, {Callaghan}, {Xue}, {Palaniappan}, {Singh}, {Antani}, {Thoma},
  {Wang}, {Lu}, and {McDonald}]{ChineseDataset2}
S.~{Jaeger}, A.~{Karargyris}, S.~{Candemir}, L.~{Folio}, J.~{Siegelman},
  F.~{Callaghan}, Z.~{Xue}, K.~{Palaniappan}, R.~K. {Singh}, S.~{Antani},
  G.~{Thoma}, Y.~X. {Wang}, P.~X. {Lu}, and C.~J. {McDonald}.
\newblock Automatic tuberculosis screening using chest radiographs.
\newblock \emph{IEEE Transactions on Medical Imaging}, 33\penalty0
  (2):\penalty0 233--245, 2014.
\newblock \doi{10.1109/TMI.2013.2284099}.
\newblock URL \url{https://doi.org/10.1109/TMI.2013.2284099}.

\bibitem[Kim et~al.(2002)Kim, Lee, Primack, Yoon, Byun, Kim, Suh, Kwon, and
  Han]{clinicalPneumonia}
E.~A. Kim, K.~S. Lee, S.~L. Primack, H.~K. Yoon, H.~S. Byun, T.~S. Kim, G.~Y.
  Suh, O.~J. Kwon, and J.~Han.
\newblock Viral pneumonias in adults: Radiologic and pathologic findings.
\newblock \emph{RadioGraphics}, 22\penalty0 (suppl\_1):\penalty0 S137--S149,
  2002.
\newblock \doi{10.1148/radiographics.22.suppl\_1.g02oc15s137}.
\newblock PMID: 12376607.

\bibitem[López-Cabrera et~al.(2021)López-Cabrera, Portal~Diaz, Orozco,
  Lovelle, and Perez-Diaz]{ShortcutCovid}
J.~López-Cabrera, J.~Portal~Diaz, R.~Orozco, O.~Lovelle, and M.~Perez-Diaz.
\newblock Current limitations to identify covid‑19 using artificial
  intelligence with chest x‑ray imaging (part ii). the shortcut learning
  problem.
\newblock \emph{Health and Technology}, 11, 10 2021.
\newblock \doi{10.1007/s12553-021-00609-8}.

\bibitem[{Maguolo} and {Nanni}(2020)]{ArtigoCritico}
G.~{Maguolo} and L.~{Nanni}.
\newblock {A Critic Evaluation of Methods for COVID-19 Automatic Detection from
  X-Ray Images}.
\newblock \emph{arXiv e-prints}, art. arXiv:2004.12823, Apr. 2020.

\bibitem[Malivenko(2018)]{py2keras}
G.~Malivenko.
\newblock pytorch2keras, 2018.
\newblock URL \url{https://github.com/nerox8664/pytorch2keras. Accessed: 01 Mar
  2021}.

\bibitem[Mercer and Salit(2021)]{diagnosis}
T.~Mercer and M.~Salit.
\newblock Testing at scale during the covid-19 pandemic.
\newblock \emph{Nature Reviews Genetics}, 22:\penalty0 1--12, 05 2021.
\newblock \doi{10.1038/s41576-021-00360-w}.

\bibitem[Montavon et~al.(2019)Montavon, Binder, Lapuschkin, Samek, and
  M{\"u}ller]{LRPBook}
G.~Montavon, A.~Binder, S.~Lapuschkin, W.~Samek, and K.-R. M{\"u}ller.
\newblock Layer-wise relevance propagation: An overview.
\newblock In \emph{Explainable AI: Interpreting, Explaining and Visualizing
  Deep Learning}, pages 193--209. Springer International Publishing, 2019.

\bibitem[Rahman et~al.(2020)Rahman, Khandakar, Qiblawey, Tahir, Kiranyaz,
  Kashem, Islam, Maadeed, Zughaier, Khan, and Chowdhury]{otherSegment}
T.~Rahman, A.~Khandakar, Y.~Qiblawey, A.~Tahir, S.~Kiranyaz, S.~B.~A. Kashem,
  M.~T. Islam, S.~A. Maadeed, S.~M. Zughaier, M.~S. Khan, and M.~E.~H.
  Chowdhury.
\newblock Exploring the effect of image enhancement techniques on covid-19
  detection using chest x-rays images, 2020.

\bibitem[Rajpurkar et~al.(2017)Rajpurkar, Irvin, Zhu, Yang, Mehta, Duan, Ding,
  Bagul, Langlotz, Shpanskaya, Lungren, and Ng]{chexnet}
P.~Rajpurkar, J.~Irvin, K.~Zhu, B.~Yang, H.~Mehta, T.~Duan, D.~Y. Ding,
  A.~Bagul, C.~Langlotz, K.~S. Shpanskaya, M.~P. Lungren, and A.~Y. Ng.
\newblock Chexnet: Radiologist-level pneumonia detection on chest x-rays with
  deep learning.
\newblock \emph{CoRR}, abs/1711.05225, 2017.
\newblock URL \url{http://arxiv.org/abs/1711.05225}.

\bibitem[Ronneberger et~al.(2015)Ronneberger, P.Fischer, and Brox]{unet}
O.~Ronneberger, P.Fischer, and T.~Brox.
\newblock U-net: Convolutional networks for biomedical image segmentation.
\newblock In \emph{Medical Image Computing and Computer-Assisted Intervention
  (MICCAI)}, volume 9351 of \emph{LNCS}, pages 234--241. Springer, 2015.

\bibitem[Sakai(2006)]{accMi}
T.~Sakai.
\newblock Evaluating evaluation metrics based on the bootstrap.
\newblock In \emph{Proceedings of the 29th Annual International ACM SIGIR
  Conference on Research and Development in Information Retrieval}, SIGIR '06,
  page 525–532, New York, NY, USA, 2006. Association for Computing Machinery.
\newblock ISBN 1595933697.
\newblock \doi{10.1145/1148170.1148261}.
\newblock URL \url{https://doi.org/10.1145/1148170.1148261}.

\bibitem[Salvatier et~al.(2016)Salvatier, Wiecki, and Fonnesbeck]{pymc3}
J.~Salvatier, T.~Wiecki, and C.~Fonnesbeck.
\newblock Probabilistic programming in python using pymc3.
\newblock 01 2016.
\newblock \doi{10.7287/PEERJ.PREPRINTS.1686V1}.
\newblock URL \url{https://doi.org/10.7287/PEERJ.PREPRINTS.1686V1}.

\bibitem[Shoeibi et~al.(2020)Shoeibi, Khodatars, Alizadehsani, Ghassemi,
  Jafari, Moridian, Khadem, Sadeghi, Hussain, Zare, Sani, Bazeli, Khozeimeh,
  Khosravi, Nahavandi, Acharya, and Shi]{reviewCovid}
A.~Shoeibi, M.~Khodatars, R.~Alizadehsani, N.~Ghassemi, M.~Jafari, P.~Moridian,
  A.~Khadem, D.~Sadeghi, S.~Hussain, A.~Zare, Z.~A. Sani, J.~Bazeli,
  F.~Khozeimeh, A.~Khosravi, S.~Nahavandi, U.~R. Acharya, and P.~Shi.
\newblock Automated detection and forecasting of covid-19 using deep learning
  techniques: A review, 2020.

\bibitem[{Stirenko} et~al.(2018){Stirenko}, {Kochura}, {Alienin}, {Rokovyi},
  {Gordienko}, {Gang}, and {Zeng}]{ShenMasks}
S.~{Stirenko}, Y.~{Kochura}, O.~{Alienin}, O.~{Rokovyi}, Y.~{Gordienko},
  P.~{Gang}, and W.~{Zeng}.
\newblock Chest x-ray analysis of tuberculosis by deep learning with
  segmentation and augmentation.
\newblock In \emph{2018 IEEE 38th International Conference on Electronics and
  Nanotechnology (ELNANO)}, pages 422--428, 2018.

\bibitem[Thomas et~al.(2019)Thomas, Heekeren, M{\"u}ller, and Samek]{fmri-lrp}
A.~W. Thomas, H.~R. Heekeren, K.-R. M{\"u}ller, and W.~Samek.
\newblock Analyzing neuroimaging data through recurrent deep learning models.
\newblock \emph{Frontiers in Neuroscience}, 13:\penalty0 1321, 2019.
\newblock \doi{10.3389/fnins.2019.01321}.
\newblock URL \url{http://dx.doi.org/10.3389/fnins.2019.01321}.

\bibitem[{Trunk}(1979)]{Curse}
G.~V. {Trunk}.
\newblock A problem of dimensionality: A simple example.
\newblock \emph{IEEE Transactions on Pattern Analysis and Machine
  Intelligence}, PAMI-1\penalty0 (3):\penalty0 306--307, 1979.

\bibitem[Wang et~al.(2020)Wang, Xu, Gao, Lu, Han, and Wu]{test}
W.~Wang, Y.~Xu, R.~Gao, R.~Lu, K.~Han, and G.~Wu.
\newblock Detection of sars-cov-2 in different types of clinical specimens.
\newblock \emph{JAMA}, 03 2020.
\newblock \doi{10.1001/jama.2020.3786}.

\bibitem[{Wang} et~al.(2017){Wang}, {Peng}, {Lu}, {Lu}, {Bagheri}, and
  {Summers}]{NIHSet}
X.~{Wang}, Y.~{Peng}, L.~{Lu}, Z.~{Lu}, M.~{Bagheri}, and R.~M. {Summers}.
\newblock Chestx-ray8: Hospital-scale chest x-ray database and benchmarks on
  weakly-supervised classification and localization of common thorax diseases.
\newblock In \emph{2017 IEEE Conference on Computer Vision and Pattern
  Recognition (CVPR)}, pages 3462--3471, 2017.
\newblock \doi{10.1109/CVPR.2017.369}.
\newblock URL \url{https://doi.org/10.1109/CVPR.2017.369}.

\bibitem[{Yang} et~al.(2018){Yang}, {Tresp}, {Wunderle}, and
  {Fasching}]{LRPMedical}
Y.~{Yang}, V.~{Tresp}, M.~{Wunderle}, and P.~A. {Fasching}.
\newblock Explaining therapy predictions with layer-wise relevance propagation
  in neural networks.
\newblock In \emph{2018 IEEE International Conference on Healthcare Informatics
  (ICHI)}, pages 152--162, 2018.
\newblock \doi{10.1109/ICHI.2018.00025}.
\newblock URL \url{https://doi.org/10.1109/ICHI.2018.00025}.

\bibitem[Zhang et~al.(2015)Zhang, Wang, and Zhao]{bayesianEstimator}
D.~Zhang, J.~Wang, and X.~Zhao.
\newblock Estimating the uncertainty of average f1 scores.
\newblock \emph{Proceedings of the 2015 International Conference on The Theory
  of Information Retrieval}, 2015.
\newblock \doi{10.1145/2808194.2809488}.

\end{thebibliography}

\end{document}